\documentclass[a4paper]{jpconf}
\usepackage{graphicx}
\usepackage{color}
\begin{document}
\title{Conformal symmetry breaking and degeneracy of
 high-lying unflavored mesons  
}

\author{Mariana Kirchbach$^1$, Adrian Pallares-Rivera$^1$, 
Cliffor Compean$^2$ and Alfredo Raya$^2$}
\address{$^1$Institute of Physics, 
Autonomous University of 
San Luis Potosi, Manuel Nava 6, San Luis Potosi, SLP 98290, Mexico\\
$^2$Institute of Physics and Mathematics,
University of  Michoacan  San Nicolas de Hidalgo,
Building C-3, University City, Morelia Michoacan 58040, Mexico
}

\ead{mariana@ifisica.uaslp.mx,pallares@ifisica.uaslp.mx,cliffor@ifm.umich.mx, 
raya@ifm.umich.mx}


\begin{abstract}
We show that though conformal symmetry can be broken by the dilaton,
such can happen without breaking the conformal degeneracy patterns in the 
spectra. Our argumentation goes as {follows:} We departure from
the gauge-gravity duality which predicts on the boundaries of the 
$AdS_5$ geometry a conformal theory, associated with QCD at high 
temperatures, and consider  ${\mathbf R}^1\times S^3$ slicing.
The inverse radius, $R$, of $S^3$ relates to the 
temperature of the deconfinement phase transition and has to satisfy,
{$\hbar c/R \gg \Lambda_{QCD}$}. 
On $S^3$, whose isometry group is $SO(4)$, we then focus on  
the eigenvalue  problem of the conformal Laplacian there, 
given by $\frac{1}{R^2}\left( {\mathcal K}^2+1\right)$, 
with $ {\mathcal K}^2$ standing for the Casimir invariant of 
the $so(4)$ algebra. This eigenvalue problem
describes the spectrum of a scalar particle, to be associated with 
a $q\bar q$ system.  Such a spectrum is characterized by a 
$(K+1)^2$-fold
degeneracy of its  levels, with $K\in [0,\infty)$. 
We then break the  conformal $S^3$ metric, 
$ds^2=d\chi^2 +\sin^2\chi (d\theta ^2 +\sin^2\theta d\varphi ^2)$ 
-in polar $\chi,\theta$, and azimuthal $\varphi$ coordinates-
according to, {$d{\widetilde s}^2=e^{-b\chi}
\left((1+b^2/4) d\chi^2 +\sin^2\chi (d\theta ^2 +
\sin^2\theta d\varphi ^2)\right)$}, and attribute the  
symmetry breaking scale $b\hbar^2c^2/R^2$ to the dilaton.
Next we show that the above metric deformation is 
equivalent to a  breaking of the conformal curvature of
 $S^3$  by a term proportional to $b\cot \chi$, 
and that the perturbed conformal Laplacian is equivalent to
$\left( \widetilde{{\mathcal K}}^2 +c_K\right)$, with $c_K$ 
a representation constant, and $\widetilde{{\mathcal K}}^2$ being 
again an $so(4)$ Casimir invariant, but this time  in 
a representation unitarily nonequivalent to the 4D rotational one.
As long as the spectra before and after the symmetry breaking 
happen to be determined each by  eigenvalues of a
Casimir invariant of an $so(4)$, no matter whether or not in a 
representation that  generates the orthogonal group $SO(4)$
as a subgroup of the conformal group $SO(2,4)$, 
the degeneracy patterns remain unaltered  though the conformal symmetry 
breaks at the level of the representation of the algebra.
We fit the $S^3$ radius and the $\hbar^2c^2b/R^2$ scale to the 
high-lying excitations in the spectra of the 
unflavored mesons, and observe the correct tendency of the  $\hbar c /R=373$ MeV value
to notably exceed $\Lambda_{QCD}$. The size of the symmetry 
breaking scale
is calculated  as $\hbar c \sqrt{b}/R=673.7$ MeV.
 \end{abstract}

\section{Introduction and background}

The five-dimensional {Anti-de Sitter} space,
$AdS_5$, is presently one of the most intriguing geometries in
field theoretical studies  due to its  relevance in establishing the link
between  brane theory and QCD. The fact is that according to the 
gauge-gravity duality conjecture \cite{Maldacena}, 
a string theory at the boundary of
$AdS_5$  appears dual to a conformal field theory in (1+3) dimensions,
associated with QCD  \cite{Witten}.
The $AdS_5$ geometry  is defined as a surface embedded in a (2+4)-dimensional
flat ambient space of two time-like ($X_0, X_5$) and four space-like 
{$(X_1,X_2,X_3, X_4$)} dimensions, 
according to \cite{Moschella}
\begin{eqnarray}
X_0^2+X_5^2-X_1^2-X^2_2-X_3^2-X_4^2&=&\nonumber\\
(X_0+X_4)(X_0-X_4)+ (X_5+X_3)(X_5-X_3) &-&(X_1-iX_2)(X_1+iX_2)=L^2,
\label{globalads5}
\end{eqnarray} 
where  $L^2$ is a scale.
In changing variables to 
\begin{eqnarray}
X_0+X_4=e^v, &\quad& X_\mu=e^vx_\mu, \quad \mu=5,1,2,3,
\label{horos}
\end{eqnarray}
the equation (\ref{globalads5}) becomes
\begin{equation}
e^v(X_0-X_4) +e^{2v}(x^2_5 -x_1^2-x_2^2-x_3^2)=L^2.
\label{warpedMink}
\end{equation}
This equation shows that the  $AdS_5$ space contains a 
flat Minkowski space-time, ${\mathbf R}^{1,3}$, warped by 
$\exp (2v)$. Dividing now by that very warp factor, and taking the
$v\to \infty$ limit,
one encounters the Minkowskian light-cone, ${\mathcal C}^{1+3}$ as,
\begin{equation}
{\mathcal C}^{1+3}:\quad \lim_{v\to \infty}
\left[ \frac{X_0-X_4}{e^v} +x^2_5 -x_1^2-x_2^2-x_3^2\right]=
\lim_{v\to \infty}\frac{L^2}{e^{2v}} \to 0.
\label{flatslice}
\end{equation} 
The $v$ variable is known as the {\it holographic\/ }
variable. Therefore, the intersections of the $AdS_5$ null-ray cone with
$X_0+X_4=e^v$ hyperplanes, produce  slices, $\Pi_v$, called branes,
which are  copies of flat Minkowski space-times 
in the $x_\mu$ coordinates with $\mu=5,1,2,3$. 
In terms of the holographic variable, 
the conformal $AdS_5$ boundary corresponds to the $v\to \infty $ limit, 
in which case the warp factor blows up. 
QCD on ${\mathcal C}^{1+3}$ as it emerges at the $AdS_5$ boundary,
is known as holographic QCD in reference to the description of confinement 
in terms of an 1D Schr\"odinger equation with a confining potential in 
the holographic dimension. A variety of interesting 
insights into the properties of hadrons have been gained within this approach
(see {ref.~\cite{BroTer} } for a recent review). 
However, ${\mathbf R}^{1,3}$, in combination with an 
interaction potential of an infinite range, represent 
a significant obstacle toward the formulation of a finite-temperature 
field theory and the description of the de-confinement phase transition.
The reason is that the absence of a scale presents a 
serious difficulty in defining the QCD chemical potential. 
This difficulty has been  circumvented \cite{Witten}, \cite{DeBoer}, 
\cite{Tommy} in replacing the flat Minkowski space-time by the 
compactified one, ${\mathbf R}^1\times S^3$, which equally well  describes  
the conformal $AdS_5$ boundary, as visible from
casting eq.~(\ref{globalads5}) into the form
{\cite{Gibbons} }
\begin{eqnarray}
\mathrm{det}U -\mathrm{ det}g=L^2, &\quad&
U=\left( 
\begin{array}{cc}
X_5+iX_0&0\\
0&X_5-iX_0
\end{array}
\right), \quad g=\left(
\begin{array}{cc}
X_4+iX_3&iX_1-X_2\\
iX_1+X_2&X_4-iX_3
\end{array}
\right),\nonumber\\
X_0+iX_5=R_0e^{i\eta}, &\quad& 
X_1+iX_2=R\sin\chi\sin\theta e^{i\varphi},\nonumber\\
X_3=R\sin\chi\cos\theta, &\quad& X_4=R\cos\chi, \quad R_0^2-R^2=L^2.
\end{eqnarray}
The proof that ${\mathbf R}^1\times S^3$ is a conformally compactified
flat Minkowski space-time, ${\mathbf R}^{1,3}$, has been  elaborated  in the 
literature in great detail \cite{Mac}, \cite{Gibbons1}, 
and the explicit map that takes the ${\mathbf R}^{1,3}$ metric to the
metric of ${\mathbf R}^1\times S^3$ can be found for example in \cite{Gibbons}.   
The $AdS_5$ line interval in global coordinates  reads, 
\begin{eqnarray}
ds^2&=& \frac{L^2}{\cos^2\alpha }
\left( -d\tau^2 + d\alpha^2 +\sin^2\alpha
\left( d\chi^2 +\sin^2\chi (d\theta^2 +\sin^2\theta d\varphi^2)\right)\right),
\label{Mink}
\end{eqnarray}
where $\alpha \in \left[0,\pi/2\right]$ parametrizes the holographic coordinate.
The boundary is approached for $\alpha \to \pi /2$ in which case 
the compactified Minkowski-spacetime, ${\mathbf R}^1\times S^3$, emerges.
Particle motions  on this space have been equally well 
studied, the ref.~\cite{Gibbons} providing valuable insights in that regard.
In particular, it has been found that the free  motion of a 
scalar particle 
is described in terms of the eigenvalue problem of
the conformal Laplacian,   ${\mathbf L}^2_{S^3}$ ,
 \begin{eqnarray}
{\mathbf L}^2_{S^3}= 
-\frac{1}{R^2}\Delta_{S^3}+{\mathcal R}_{S^3} , &\quad&
- \frac{1}{R^2}\Delta_{S^3}= -
\frac{1}{R^2}
\left( \frac{{\mathrm d}^2}{{\mathrm d}\chi ^2} 
+2\cot\chi 
\frac{{\mathrm d}}{{\mathrm d}\chi} -
\frac{{\mathbf L}^2}{\sin^2\chi}\right)=\frac{1}{R^2}{\mathcal K}^2,
\label{LapBelt}
\end{eqnarray}
where $\Delta_{S^3}$ is the Laplace-Beltrami operator on $S^3$,
${\mathcal R}_{S^3}$ stands for the conformal curvature, and
${\mathcal K}^2$ is the operator of the squared four-dimensional 
(4D) angular momentum which acts as a Casimir invariant of the isometry algebra
$so(4)$ of $S^3$. The $so(4)$ algebra in
the representation chosen in (\ref{LapBelt}) has the property
to  generate the orthogonal group 
$SO(4)$. On the unit hypersphere, the spectral problem of 
${\mathbf L}_{S^3}$  reads,
\begin{eqnarray}
{{\mathbf L}_{S^3}Y_{Klm}(\vec\Omega)}=
(-\Delta_{S^3} +1)Y_{Klm}(\vec\Omega)&=&
\left( {\mathcal K}^2+1\right)Y_{Klm}(\vec\Omega)=
( K+1)^2Y_{Klm}(\vec\Omega),\quad \vec \Omega=({\chi,\theta,\varphi})
\nonumber\\
{ Y_{Klm}(\vec\Omega)}= {\mathcal S}_K^l(\chi)
Y_l^m(\theta, \varphi), &\quad&
 {\mathcal S}_K^l(\chi)=\sin^l\chi {\mathcal G}_{K-l}^{l+1}(\cos \chi), 
\nonumber\\
\quad K\in[0,\infty), \quad l\in \lbrack 0, K \rbrack,
&\quad& m\in \lbrack -l,+l\rbrack .
\label{four_AM}
\end{eqnarray}
Here, {$Y_{Klm}(\vec\Omega)$} are the hyper-spherical harmonics,
${\mathcal G}_n^\alpha(\cos \chi)$ denote the Gegenbauer polynomials, and
$Y_l^m(\theta, \varphi)$ stand for the ordinary three-dimensional
spherical harmonics. The $\chi$ dependent part of the
hyper-spherical harmonics, i.e. the ${\mathcal S}_K^l(\chi)$ function,
is sometimes referred to as ``quasi-radial'' function {\cite{Vinitsky}}, 
a notation which we occasionally  shall make  {use of}.
It is obvious, that the spectrum in (\ref{four_AM}) 
is characterized by a $(K+1)^2$ fold 
degeneracy of the levels. Such a spectrum as whole  would fit into 
an infinite unitary
representation of the conformal group, $SO(2,4)$, where the group $SO(4)$
appears in the reduction chain,
\begin{equation}
SO(2,4)\subset SO(4)\subset SO(3)\subset SO(2).
\label{redchain}
\end{equation}
{It} is within this context that  $(K+1)^2$-fold degeneracy is 
associated with conformal symmetry, in parallel to the H atom where
$(K+1)$ acquires  the meaning of the principal quantum number of the 
Coulomb potential  problem. 
However, conformal symmetry can be at most an approximate symmetry
of QCD because it requires 
\begin{itemize}
\item massless quarks, 
\item scale independent  
strong coupling, and 
\item  a massless dilaton.
\end{itemize}
 While the first two conditions seem to be reasonably 
respected within the unflavored sector
of QCD, in which the $u$ and $d$ quark masses can be considered 
sufficiently small at the scale of the nucleon excitations, which start 
around 1500 MeV, and in the infrared regime, where the running 
coupling constant 
starts walking  toward a fixed value {\cite{Andre}}, the dilaton mass 
can  not be disregarded so far. The breaking of the conformal
symmetry is frequently treated in terms of
breaking the metric of the Minkowski space-time encountered
at the $AdS_5$ boundary, i.e.  by deforming the warp factor
(soft-walls) \cite{BroTer}. Notice, 
the dilaton can be introduced in the action, ${\mathcal S}$, 
as a background field in an overall exponential according to, 
\begin{eqnarray}
{\mathcal S}=\int {\mathrm d}^{D}Xe^{-\Phi (z)}\sqrt{-g}{\mathcal L}, \quad \Phi (z)=
(\mu z)^\nu , \quad z=e^{-v},
\label{dilaton_scale}
\end{eqnarray}
where $\mu $ sets a mass scale, usually taken as $\Lambda_{QCD}$ for simplicity.
As shown in \cite{Colangelo}, \cite{Kelley}, the dilaton 
modifies the particle motion and  leaves a print in the meson spectra.

\begin{quote}
We here adapt the essentials of the above scheme to the geometry of the 
compactified Minkowski space-time, where the holographic dimension can be kept for the time being
as a spectator insofar as particle motion on the finite volume $S^3$ space is automatically confined.
We design  a pragmatic quark model of conformal symmetry breaking on ${\mathbf R}^1\times S^3$ through 
global $S^3$ metric deformation by an exponential
factor of the second polar angle. We then
study the impact of this deformation   on the spectra
of the high-lying unflavored mesons reported  
by the Cristal Barrel collaboration as incorporated in the compilation of {\cite{Afonin_SS}}.
Our aim  is to reveal  a possibility to deform 
the conformal curvature in (\ref{Mink})  by a scale, 
and without removing the conformal degeneracy patterns from the spectra.
\end{quote}

The outline of the paper is as follows.
In the next section we present the mechanism of conformal symmetry breaking
without removing the conformal degeneracy patterns in the spectra.
In section 3 we analyze the  data  on
the high-lying unflavored meson spectra listed in ~{\cite{Afonin_SS}} with the aim
to  obtain an estimate for the $S^3$ radius and the order of magnitude of
the conformal symmetry breaking scale which we then attribute to the dilaton. 
The paper ends with concise conclusions.

\section{Breaking the conformal curvature  of  $S^3$ 
by an interaction}

Any perturbation of the quantum mechanical motion on $S^3$ by an interaction potential, 
$V({\vec\Omega})$, breaks  the conformal curvature as,
\begin{equation}
\hbar^2c^2{\mathcal R}_{S^3}\longrightarrow 
\hbar^2c^2 \widetilde{{\mathcal R}}_{S^3}= 
\frac{\hbar^2c^2}{R^2}+{V(\vec\Omega)},
\label{broken_metric}
\end{equation} 
and the perturbed Laplacian is no longer conformal, neither is its 
spectral problem necessarily exactly solvable.
Not so for the potential, given by,
\begin{equation}
V({\chi })=-2B\cot\chi,
\label{Curved_Coulomb}
\end{equation}
with $B=b\hbar^2c^2/R^2$, and a dimensionless $b$, 
which gives rise to an exactly solvable $\left( 
\hbar^2c^2{\mathbf L}_{S^3}+V({\chi })\right)$
spectral problem, and one which moreover
happens to carry same   $(K+1)^2$-fold
degeneracy patterns  as  the spectral problem of the
intact conformal Laplacian in (\ref{four_AM}).
The $\cot\chi$ potential is special insofar as it is a harmonic function 
on $S^3$, just as is the Coulomb potential in $E_3$, and can be 
treated along the line of harmonic analysis and potential theory.
Using harmonic functions as potentials brings the advantage  of minimizing the
Dirichlet functional integral.
Occasionally, the cotangent potential on $S^3$ is referred to as 
``curved'' Coulomb potential.
In the present section we wish to take a closer look on the quantum mechanical
cotangent potential problem on $S^3$ from 
the perspective of conformal Laplacians {\cite{Rosenberg,Matthew}}. 
The cotangent potential  problem on $S^3$, first considered by
Schr\"odinger {\cite{Schr40,Schr41}}, is equivalent to
the relativistic  four-dimensional rigid-rotator problem perturbed by a 
$\cot\chi$ interaction,

\begin{eqnarray}
\frac{{\hbar^2}c^2}{R^2}\left[-\Delta_{S^3}+1 -
2b\cot\chi\right]{\Psi_{K{\widetilde l}{\widetilde m}}(\vec\Omega)}&=&
 \frac{\hbar^2c^2}{R^2}\left[(K+1)^2 -\frac{b^2}{(K+1)^2}\right]
{\Psi_{K{\widetilde l}{\widetilde m}}(\vec\Omega)}.
\label{RM_1}
\end{eqnarray}
The solutions of equation (\ref{RM_1}) are well studied and known. 
Specifically in  {\cite{JPhA_2011}}, they have been expressed
in terms of real
non-classical Romanovski polynomials, $R_n^{\alpha, \beta}(\cot\chi)$
(reviewed in {\cite{raposo}}),  and in units of $\hbar=1=c$, $R=1$, read,
\begin{eqnarray}
{\Psi_{K{\widetilde l}{\widetilde m}}(\vec\Omega)}&=& 
e^{\frac{\alpha_K\chi}{2}}\psi_K^{\widetilde l}
(\chi)Y_{\widetilde l}^{\widetilde m}(\theta, \varphi),\nonumber\\
\psi_K^{\widetilde l}(\chi)&=&\sin^K\chi R_{K-{\widetilde l}}^{\alpha_K,\beta_K-1}
(\cot\chi) ,\nonumber\\
\alpha_K=-\frac{2b}{K+1}, &\quad& \beta_K=-K.
\label{RM_2}
\end{eqnarray}

The Romanovski polynomials 
satisfy the following differential {hyper-geometric} equation,
\begin{equation}
(1+x^2)\frac{{\mathrm d}^2R_n^{\alpha, \beta}}{{\mathrm d} x^2}
+2\left(\frac{\alpha }{2} +\beta x
\right)\frac{{\mathrm d}R_n^{\alpha, \beta}}{{\mathrm d}x}
-n(2\beta +n-1)R_n^{\alpha, \beta}=0.
\end{equation}
They are obtained from the following  weight function,
\begin{equation}
\omega ^{\alpha, \beta}(x)=(1+x^2)^{\beta -1}\exp(-\alpha \cot^{-1}x),
\end{equation}
by means of the Rodrigues formula,
\begin{equation}
R^{\alpha, \beta}_n(x)=\frac{1}{\omega^{\alpha, \beta }(x)}
\frac{{\mathrm d}^n}{{\mathrm d}x^n}
\left[ (1+x^2)^n\omega^{\alpha, \beta}(x)\right].
\end{equation}
Recently, it has been shown in \cite{JPhA_2011} that the
$\psi_K^{\widetilde l} (\chi )$ functions allow for finite decompositions in 
the $SO(4)$ basis of the quasi-radial  functions $S_K^l(\chi)$ in
(\ref{four_AM}).
Namely, the following relation has been found to {hold valid,} 
\begin{eqnarray}
\psi_{K}^{\widetilde l} (\chi )={
\sum_{{\widetilde l}}^{K}C_{ l}\ {\mathcal S}_K^{l}(\chi).}
\label{Gegenbauer}
\end{eqnarray}
The explicit expressions for the coefficients {$C_{l}$} defining
the full  similarity transformation between the  
invariant spaces of the perturbed and intact Laplacians
on $S^3$ can be consulted  in \cite{JPhA_2011}. 
Upon introducing the notations, 
\begin{equation}
\omega_K(\chi)=\frac{\alpha_K\chi}{2}{\equiv-\frac{b\chi}{K+1}},
\qquad  b=-\omega^\prime_K (\chi)(K+1),
\label{omega_alpha_K}
\end{equation}
and insertion of (\ref{Gegenbauer}) into (\ref{RM_2}),  allows to rewrite
(\ref{RM_1}) to,
\begin{eqnarray}
\left[{\bf L}_{S^3}+
2\omega^\prime_K(\chi)(K+1) \cot\chi\right]e^{\omega_K(\chi )}
\psi_K^{\widetilde l}(\chi)&=&\nonumber\\
&&\hspace{-8mm}{
e^{\omega_K(\chi )}\sum_{\widetilde l}^KC_{ l} \left[ 
(K+1)^2  -\omega^\prime_K (\chi)\, ^2\right]S_K^{ l}(\chi).}
\label{RM_3}
\end{eqnarray}
Dragging now the exponential factor in the l.h.s. in (\ref{RM_3})
from the right to the very left, recalling that
${\mathbf L}_{S^3}={\mathcal K}^2+1$, and making use of (\ref{Gegenbauer}), 
 amounts to
\begin{eqnarray}
e^{\omega _K(\chi )}
\left[
{\mathcal  K}^2 +1  - \omega_K^\prime (\chi)\, ^2
-2\omega^{\prime} _K(\chi ){\mathcal D}_{\mathbf K}
\right]{\sum_{\widetilde l}^{K}C_{ l}\ {\mathcal S}_K^{ l}(\chi)}
&=&\nonumber\\
&&\hspace{-22mm}{
e^{\omega_K(\chi )}\sum_{\widetilde l}^KC_{l} \left[ 
(K+1)^2  -\omega^\prime_K (\chi)\, ^2\right]S_K^{ l}(\chi),} \nonumber\\
{\mathcal D}_{\mathbf K}&=&\frac{{\mathrm d}}{{\mathrm d}\chi}-K\cot \chi.
\label{boa_boa_2}
\end{eqnarray}
Using in the r.h.s. of (\ref{boa_boa_2}) the relationship,  
$\left( {\mathcal K}^2+1\right)S_K^{ l}(\chi)=(K+1)^2S_K^{ l}(\chi)$, known 
from (\ref{LapBelt}), allows to equivalently cast eq.~(\ref{boa_boa_2}) as,
\begin{eqnarray}
{\Big(}
\widetilde{{\mathcal  K}}^2+1  - 
\omega_K^\prime (\chi)\, ^2
&-&2\omega^{\prime} _K(\chi )e^{\omega _K(\chi )}{\mathcal D}_{\mathbf K}
e^{-\omega _K(\chi )}
{\Big)}
e^{\omega _K(\chi )}
{
\sum_{\widetilde l}^{K}C_{ l}\ {\mathcal S}_K^{l}(\chi)}\nonumber\\
&&\hspace{-14mm}
= \ 
\sum_{\widetilde l}^{K} \left(
\widetilde{ {\mathcal K}}^2 \,
+1 -\omega^\prime(\chi)^2\right)  
e^{\omega_K(\chi)}
{  C_{ l}\ {\mathcal S}_K^{ l}(\chi).}
\label{hu_hu}
\end{eqnarray}
This equality implies the following important property of the 
${\mathcal D}_K$ operator,

\begin{eqnarray}
{\Big(}
\widetilde{{\mathcal  K}}^2   
-2\omega^{\prime} _K(\chi )e^{\omega _K(\chi )}{\mathcal D}_{\mathbf K}
e^{-\omega _K(\chi )}
{\Big)}
e^{\omega _K(\chi )}
{\sum_{\widetilde l}^{K}C_{ l}\ {\mathcal S}_K^{ l}(\chi)}
&=& 
\sum_{\widetilde l}^{K}\widetilde{\mathcal K}^2  
\,e^{\omega_K(\chi)} 
{C_{l} \ {\mathcal S}_K^{l}(\chi),}\nonumber\\
\widetilde{\mathcal K}^2&=&e^{\omega_K(\chi)}{\mathcal K}^2 e^{-\omega_K(\chi)}.
\label{husch_husch}
\end{eqnarray}
Now, the equivalence of eqs.~(\ref{RM_3})--(\ref{boa_boa_2}), 
and (\ref{hu_hu})  allows to draw the final conclusion as
\begin{eqnarray}
\left[{\mathcal K}^2+
1+2\omega^\prime_K(\chi)(K+1) \cot\chi\right]e^{\omega(\chi)}
\psi_K^{\widetilde l}(\chi)&=&
{\sum_{\widetilde l}^{K} C_{ l} \left[\widetilde{\mathcal K}^2 +
1 -\omega^\prime (\chi)^2\right] \,e^{\omega_K(\chi)}{\mathcal S}_K^{ l}(\chi)}\nonumber\\
&=& \left[ (K+1)^2 -\omega^\prime_K (\chi)\, ^2\right]
e^{\omega(\chi)}\psi_K^{\widetilde l}(\chi),
\label{that_is}
\end{eqnarray}
with $\psi_K^{\widetilde l}$ from (\ref{Gegenbauer}), 
and $\omega_K^\prime (\chi)$ being a representation constant.
In this fashion, the eigenvalue problem of the cotangent-broken
conformal Laplacian on $S^3$ presents itself as the eigenvalue problem of
a Casimir invariant of the $so(4)$ algebra in a representation unitarily 
nonequivalent to the 4D-rotational.
As long as the eigenvalues of any Casimir invariant remain 
unaltered under any well defined similarity transformations, 
no matter unitary or not, the degeneracy patterns in the spectral problems of 
the conformal  and the   cotangent-broken Lapalacians result same. 
The Killing vectors of this new $so(4)$ algebra,
\begin{eqnarray}
 \widetilde{J}_{ik}=e^{\omega_K (\chi)}J_{ik}e^{-\omega_K(\chi)}&=&
e^{\omega_K(\chi)}
\left( X_j\frac{\partial}{\partial X_k} -
X_k\frac{\partial}{\partial X_j} 
\right)e^{-\omega_K(\chi)},\nonumber\\
\left[\widetilde{ J}_{ik},\widetilde{ J}_{kj}  \right]=
\widetilde{ J}_{ji}, &\quad&
\chi=\cot^{-1}\frac{X_4}{\sqrt{X_1^2+X_2^2+X_3^2}}\;,
\label{Killing_S3}
\end{eqnarray}
refer to a surface (call it ${\widetilde S}^3$) which will  differ in 
shape  from $S^3$.
In this fashion, the breaking of the conformal symmetry reveals itself 
through metric deformation. Equivalently, one can say that it reveals 
itself at the level of the representation functions, in reference to 
the difference between
the hyper-spherical harmonics,
{$Y_{Klm}(\vec\Omega)$}
in (\ref{four_AM}), and  $\Psi_{K\widetilde{l}\widetilde{m}}(\vec\Omega)$
in (\ref{RM_2}). 
Indeed, the global metric of the $S^3$ ball is incorporated by 
the scalar $so(4)$ representation, which, in being
the lowest ${\mathbf L}_{S^3}$ eigenstate,  is the hyper-spherical harmonic, 
{$Y_{000}(\vec\Omega)$}, 
\begin{eqnarray}
S^3:\quad {Y_{000}(\vec\Omega)}&=&
\sum_{i=1}^{4}X_i^2=
\cos^2 \chi +\sin^2\chi\left( \cos^2\theta +\sin^2\theta 
\left( 
\cos^2\varphi +\sin^2\varphi 
\right) 
\right)=1.
\label{S3_global_mtrc}
\end{eqnarray}
Correspondingly, the global metric of the ${\widetilde S}^3$ surface 
will be implemented by the scalar of the non-unitary $so(4)$ algebra 
in (\ref{husch_husch}), (\ref{Killing_S3}) and is given by the exponentially rescaled hyper-sphere,
\begin{eqnarray}
{\widetilde S}^3:\quad e^{-b\chi }{Y_{000}(\vec\Omega)} \ =\
e^{-b\chi}\left( \cos^2 \chi +\sin^2\chi\left( \cos^2\theta +\sin^2\theta 
\left( 
\cos^2\varphi +\sin^2\varphi 
\right) 
\right)\right)=e^{-b\chi}.
\label{S3_global_deformed}
\end{eqnarray}
The line element on the deformed surface  
${\widetilde S}^3$ is then easily calculated as,

\begin{eqnarray}
{\mathrm d}^2 {\widetilde s}&=&
e^{-b\chi}\left( 
(1+b^2/4) {\mathrm d}^2\chi +
\sin^2\chi \left( {\mathrm d}\theta^2 +\sin^2\theta {\mathrm d}^2\varphi
\right)\right).
\label{def_me}
\end{eqnarray}
In the following we conjecture that the $\hbar^2c^2b/R^2$ scale
of conformal symmetry breaking can be attributed  to the dilaton, the only 
significant source of such a breaking.

Notice  that the perturbance of the conformal Laplacian
on $S^3$ by a cotangent interaction does not even produce  a conformal map 
\cite{Rosenberg}, \cite{Matthew},
as it could have happened by a different, in general, gradient potential.

\begin{quote}
In summary, a subtle mode of symmetry breaking has been identified
in which the symmetry breaking happens at the level of the representation 
functions of the isometry algebra of a given geometric manifold, and  
reveals itself through a metric deformation, while the spectral problems of the
intact and  broken Laplacians are characterized by identical
degeneracy patterns.
\end{quote}
 It seems opportune to refer to this symmetry breaking mechanism as
``symmetry breaking camouflaged  by degeneracy'', or, shortly, 
``camouflaged  symmetry breaking''.

\section{Degeneracy of high-lying unflavored mesons}
In the following we  employ the
cotangent-broken conformal Laplacian on $S^3$ in (\ref{RM_1})
in the  description of the  spectra of the high-lying 
unflavored mesons as  compilation in {\cite{Afonin_SS}} and adjust the 
$R$ and $b$ potential parameters.  
The cotangent potential derives its utility for employment in the 
description of hadron spectra from its close  relationship to 
the Cornell potential,  predicted by Lattice QCD {\cite{Cornl}},
as visible from  its Taylor series decomposition,
\begin{equation}
-2b\cot \frac{\stackrel{\frown}{r}}{R}
= -\frac{2bR}{\stackrel{\frown}{r}}
+\frac{4b}{3R}\stackrel{\frown}{r}+..., \quad \mbox{for}\quad
\chi=\frac{\stackrel{\frown}{r}}{R},
\label{Cornell}
\end{equation}
where $\stackrel{\frown}{r}$ denotes the geodesic distance on $S^3$, and $R$
was the $S^3$ radius.

The  data on the spectra of the high-lying unflavored mesons, and specifically the Crystal Barrel data, 
show well pronounced conformal degeneracy patterns in the region above 
$\approx$ 1200 MeV.  Our case is that 
\begin{itemize}
\item
these patterns emerge in consequence of
the breaking of the conformal $SO(2,4)$ group symmetry at the level
of the representation of its algebra, and
 in accordance with eqs.~(\ref{that_is}), (\ref{def_me}),
\item  such a breaking may be attributed to the dilaton.

\end{itemize}
 
In Figs.~(1--4) we present the spectra of the mesons of interest.
Tables  1 and 2 contain the results of the {least} square parameter fit.

\begin{center}
\begin{table}[h]
\caption{The potential parameters from the
least square fit to the meson spectra by means of eq.~(\ref{RM_1}).}
\centering
\begin{tabular}{cccc}
\br
Isospin&Lowest spin& R [fm]&($\hbar^2c^2b)/R^2$[GeV$^2$] \\
\mr
I=0& $J_{\textup{\footnotesize min}}$=0$^-$  & 0.5157279 & 0.459666    \\
I=0& $J_{\textup{\footnotesize min}}$=1$^-$  & 0.5293829 & 0.425489    \\
I=1& $J_{\textup{\footnotesize min}}$=0$^-$  & 0.5443879 & 0.522783   \\
I=1& $J_{\textup{\footnotesize min}}$=1$^-$  & 0.5268850 & 0.411833   \\
\br
\end{tabular}
\end{table}
\end{center}

\begin{center}
\begin{table}[h]
\caption{The meson masses, $M^{\mbox{\footnotesize fit}}$, from the least square fit to the 
experimental values, $M^{\mbox{\footnotesize expr}}$,
by means of eq.~(\ref{RM_1}). The label $K$ stands for the
four-dimensional angular momentum characterizing a degenerate level in accordance with eq.~(\ref{RM_1}). The fourth column contains the averaged mass of the
experimentally observed states as displayed in the respective plots on 
the Figs.~1-4.}
\centering
\begin{tabular}{ccccl}
\br
I  & J$_{\min}$ & K&  $M^{\mbox{\footnotesize expr} }$[MeV] &$M^{\mbox{\footnotesize fit} }$ [MeV] \\
\mr
0& 0$^-$   & 1 &  1278.33  & 1349.60 \\
   &           & 2 &  1692.5   & 1659.48\\
   &           & 3 &  1970.66  & 1961.83\\
   &           & 4 &  2274.22  & 2280.1 \\
1& 0$^-$   & 1 &  1328.33  & 1404.89 \\
   &           & 2 &  1696.66  & 1708.67\\
   &           & 3 &  2031     & 1985.04\\
   &           & 4 &  2253.75  & 2273.69\\
0& 1$^-$   & 1 &  1325     & 1416.19   \\
   &           & 2 &  1653.33  & 1697.43\\
   &           & 3 &  1994.14  & 1979.19\\
   &           & 4 &  2269.77  & 2279.69\\
1& 1$^-$   & 1 &  1343.33  & 1385.85 \\
   &           & 2 & 1682.5    &1670.28\\
   &           & 3 & 1825.85   &1957.16\\
   &           &4  & 2256.66   & 2262.75\\
\br
\end{tabular}
\end{table}
\end{center}

\begin{figure}
\begin{center}
\includegraphics{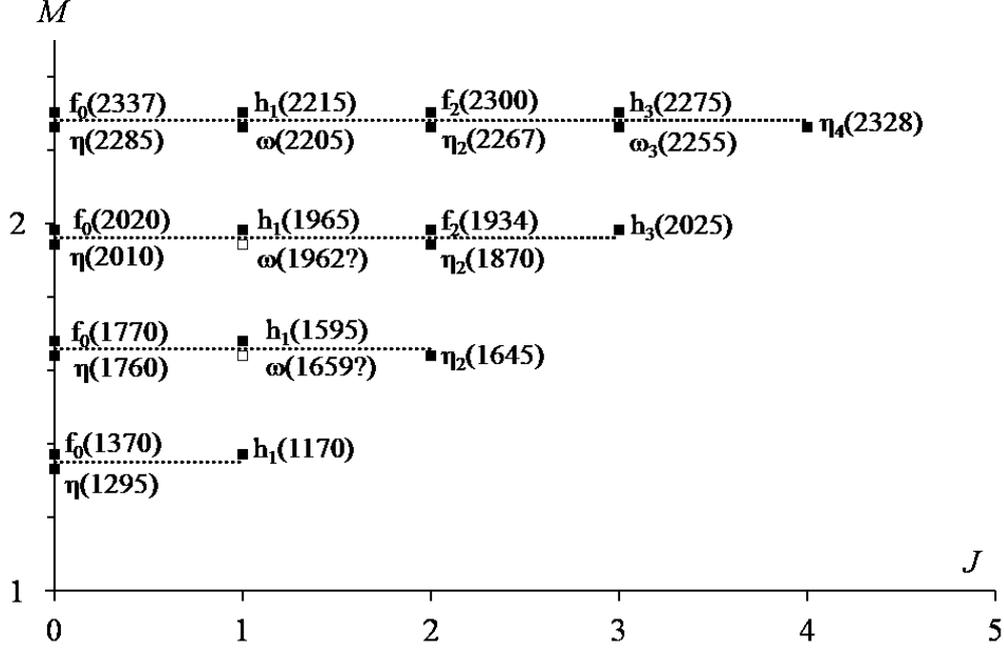}
\end{center}
\caption{High-lying excitation masses, $M$ [GeV],  
of the isosinglet pseudoscalar
$\eta $  meson according
the data  compilation of \cite{Afonin_SS}.
On the one side, the quantum numbers of the mesons in the
four groups displayed here fit well into the respective $K=1,2,3$, and $4$ 
eigenstates of the conformal Laplacian on $S^3$ in ~(\ref{four_AM}).
At the same time, they  fit equally well  into the
eigenstates of  the broken  Laplacian in (\ref{that_is}), which shows that
the breaking of the conformal group symmetry at the level of 
the representation of its algebra by the $\hbar^2c^2b/R^2$ mass scale  
does not leave a print
in the degeneracies in the spectrum.
Empty squares denote ``missing'' resonances. These are additionally
marked by a question-mark inside the round brackets, while the
accompanying number is our  prediction for the respective mass.  
Notice that not only the conformal degeneracy patterns are well
pronounced, also practically all the states with spins lower than the maximal,
in a level, $(J_{\textup{\footnotesize max}}=K)$,  appear parity duplicated, 
a phenomenon that is indicative of a manifest realization of the chiral symmetry in the 
Wigner-Wyle mode at that scale, or, remains of it after a moderate chiral symmetry breaking.   }
\end{figure}

\begin{figure}
\begin{center}
\includegraphics{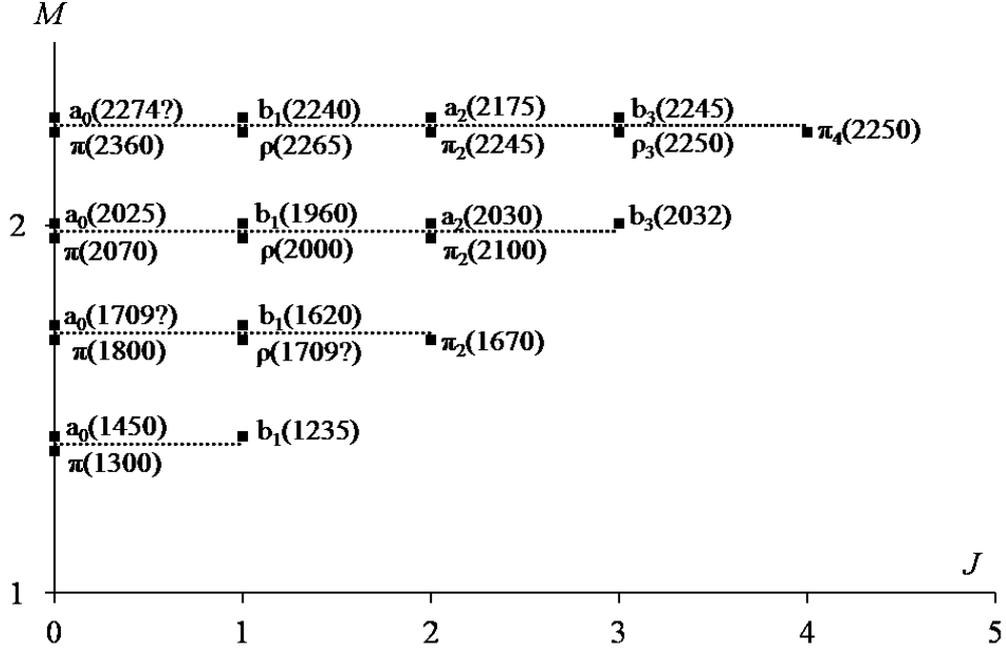}
\end{center}
\caption{High-lying excitations of the isotriplet  pseudoscalar 
$\pi$ meson according  to  
the data compilation of \cite{Afonin_SS}.
For notations and clarifying  comments see Fig.~1.}
\end{figure}

\begin{figure}
\begin{center}
\includegraphics{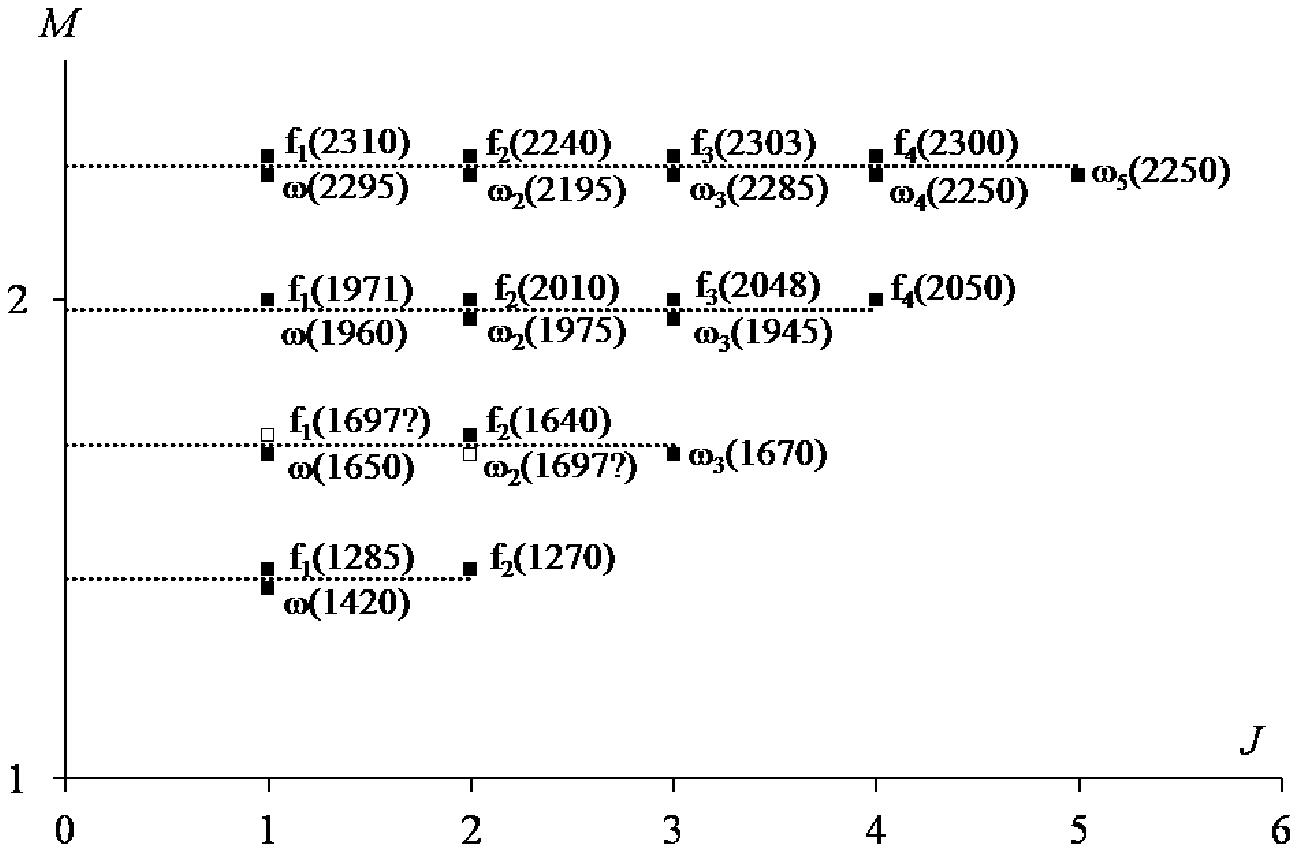}
\end{center}
\caption{High-lying excitations of the isosinglet vector 
$\omega $ meson according to the data 
compilation of \cite{Afonin_SS}. For notations and clarifying  comments see Fig.~1.}
\end{figure}

\begin{figure}
\begin{center}
\includegraphics{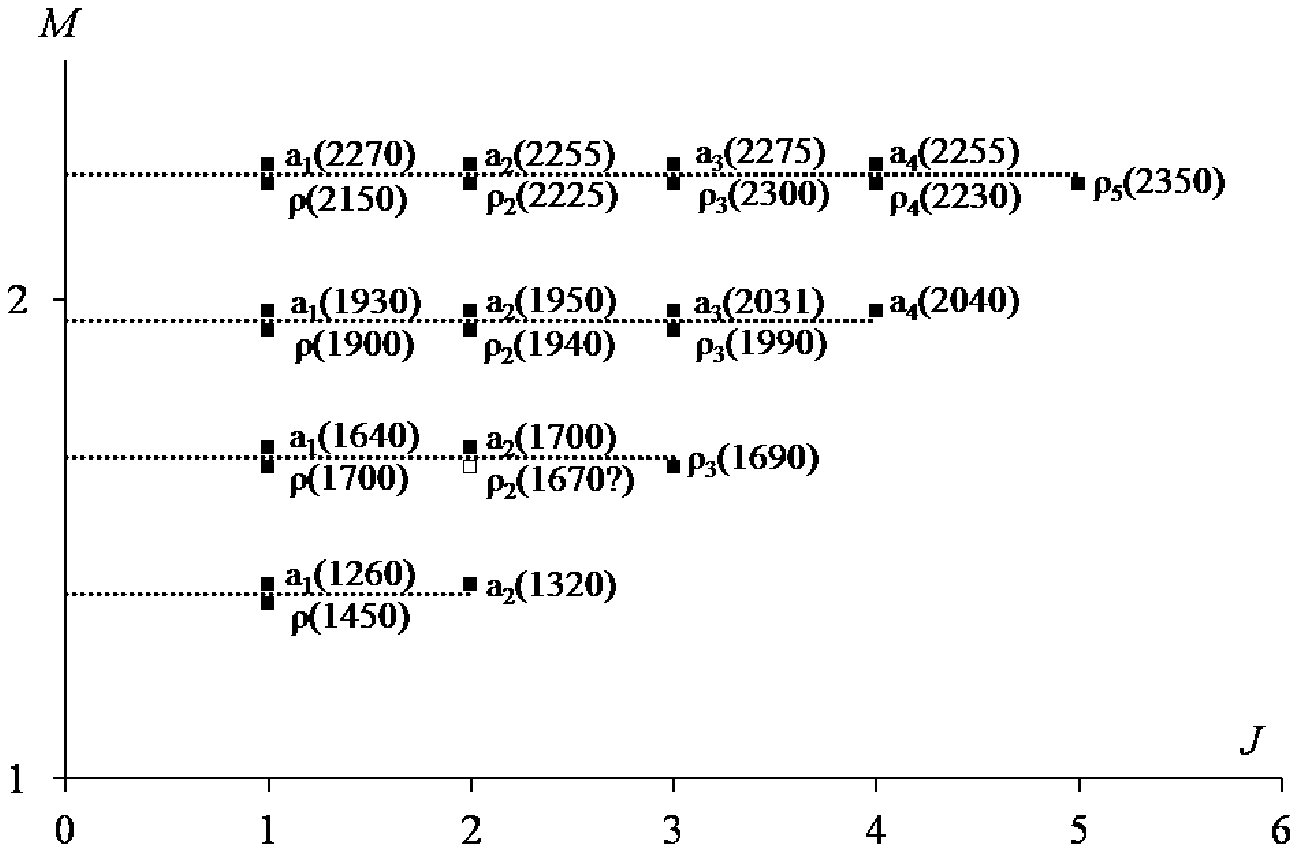}
\end{center}
\caption{High-lying excitations of the isotriplet vector  
$\rho $ meson according to the data compilation of 
\cite{Afonin_SS}. For notations and clarifying  comments see Fig.~1.}
\end{figure}

\section{Discussion and conclusions}

The Tables  1 and 2 show that our least square fits to   
on the data on the masses of the high-lying unflavored mesons by means of the cotangent broken conformal Laplacian 
on $S^3$ in eq.~(\ref{RM_1}) provide quite a reasonable description of the
observed degeneracies in the spectra under investigation and in support
of our hypothesis that the breaking of the conformal symmetry 
can occur at the level of the representation of 
the conformal algebra  and without affecting the degeneracies.
The $b$ parameter is obtained as
\begin{eqnarray}
b&=&3.2793\pm 0.0697.
\label{b-error_bars}
\end{eqnarray}
We furthermore observe that to a very good approximation, 
both the $S^3$ radius and the $\hbar^2c^2b/R^2$ scale 
are isospin and parity independent, as should be for a 
conformal symmetry breaking  scale due to the dilaton. 
The mean value of the inverse radius obtained from our fits
corresponds to a temperature of  $T=\hbar c/R$= 373 MeV which  
reasonably fulfills the requirement to be notably larger than 
$\Lambda_{QCD}=175$ MeV.
In order to extract the scale parameter of the conformal symmetry 
breaking within our approach, we make use of
eq.~(\ref{Cornell}), i.e. of  $\chi=\frac{\stackrel{\frown}{r}}{R}$, and 
rewrite the $\exp (-b\chi)$ factor equivalently as, 
\begin{eqnarray}
\exp (-b\chi) &=&\exp \left(-\left[
\left( 
\frac{\hbar c\sqrt{ b}}{R}
\right)^2
\left(\frac{\sqrt{\stackrel{\frown}{r} R}}{\hbar c}\right)^2
\right]
\right)\equiv \exp(-\mu^2 r^2),\nonumber\\
\mu=\frac{\hbar c \sqrt{ b}}{R}, &\quad& r=\frac{\sqrt{\stackrel{\frown}{r}  R}}{\hbar c}.
\end{eqnarray}
Evaluating this scale with the  numbers listed in Table 1 amounts to a 
mean value of $\mu =673.7 $ MeV, which is closer to the dilaton mass than to 
$\Lambda _{QCD}=175$ MeV in (\ref{dilaton_scale}). 
We conclude that the data on 
high-lying unflavored mesons support our model of conformal symmetry breaking
by a mass scale at the level of the representation of the 
underlying $so(4)$  algebra,
and without loosing the degeneracies in the spectra.


\newpage


\section*{References}

\begin{thebibliography}{99}

\bibitem{Maldacena} Maldacena J 1998,
{\em Phys.\ Rev.\ Lett.}\ {\bf 80}, 4859.

\bibitem{Witten} Witten E 1998,
{\em Adv.\ Theor.\ Math.\ Phys.}\ {\bf 2}, 233.

\bibitem{Moschella} Moschella U 2005, 
{\em S\'eminaire Poincar\'e} {\bf 1}, 1.

\bibitem{BroTer}  de Teramond G F and Brodsky S J 2011,
{\em J.\ Phys.\ Conf.\ Ser.} {\bf 287}, 012007.


\bibitem{DeBoer} de Boer J 2002, 
{\em  Proceedings  of the 10 Int. Conf. on Supersymmetry and Unification of 
 Fund. Interactions (SUSY02), Hamburg, Vol.} {\bf  1}, 512.

\bibitem{Tommy} Hands S, Hollowood T J and  Myers J C 2010,
{\em JHEP} {\bf 1007} 086.

\bibitem{Gibbons}Gibbons G W and Steif A R 1995,
{\em Phys.\ Lett.\ B} {\bf 346}, 255.

\bibitem{Mac} L\"uscher M and  Mack G 1975,
 {\em Commun.\ Math.\ Phys.}\ {\bf 41}, 203.

\bibitem{Gibbons1} Gibbons G W 2011  Anti-de-Sitter space and its uses,
{\em Preprint} arxiv:1110.1206v1 [hep-th].

\bibitem{Vinitsky} Pervushin V N, Pogosyan G S, 
 Sissakian A N and  Vinitsky S I 1993,
{\em Phys.\ Atom.\ Nucl.} {\bf 5b}, 1027.

\bibitem{Andre} Deur A, Burkert V,  Chen J P and  Korsch W 2008,
{\em Phys.\ Lett.} {\bf 665}, 349.

\bibitem{Colangelo}  Colangelo P,  De Fazio F,  Gianuzzi F, 
 Jugeau F and Nicotri S 2008, 
{\em Phys.\ Rev.\ D} {\bf 78}, 055009.
\bibitem{Kelley} Kelley T M, Bartz S B and  Kapusta J I,
{\em Phys.\ Rev.\ D} {\bf 83}, 016002 (2011).
\bibitem{Afonin_SS}  Afonin S S 2006,
{\em Eur.\ Phys.\ J.\ A} {\bf 29}, 327;
Afonin S S 2008,
{\em Int.\ J.\ Mod.\ Phys.\  A} {\bf 23}, 4205.
 \bibitem{Rosenberg} Rosenberg S 1998, 
{\it The Laplacian on a Riemannian manifold\/}
(London: London Mathematical Society, Student Text 31, 
Cambridge Univerisity Press).

\bibitem{Matthew} Gurski M J 2006,
{\it Conformal invariants and non-linear elliptic equations,\/}
{\em Proc. of the Int. Congress of Mathematicians,
Madrid, Spain, pp. 203-212.}

\bibitem{Schr40}  Schr\"odinger E 1940,
{\em Proc.\ Roy.\ Irish Acad.\ A} {\bf 46}, 9.

\bibitem{Schr41} Schr\"odinger E 1941,
{\em Proc.\ Roy.\ Irish Acad.\,} {\bf 46}, 183.

\bibitem{JPhA_2011} 
Pallares-Rivera A and Kirchbach M 2011, 
{\em J.\ Phys.\ A:Math.Theor.} {\bf 44}, 445302.


\bibitem{raposo} Raposo A, Weber H J,  Alvarez-Castillo D E and
Kirchbach M 2007, 
{\em C.\ Eur.\ J.\ Phys.} {\bf 5}, 253.


\bibitem{Cornl} Takahashi T T,  Suganuma H,  Nemoto Y and 
Matsufuru H 2002, 
{\em Phys.\ Rev.\ D} {\bf 65}, 114509.

\end{thebibliography}
\end{document}